\pdfoutput=1
\documentclass[12pt,letterpaper,floatfix]{article}
\usepackage[super,compress]{cite}
\usepackage{graphicx}
\usepackage[utf8]{inputenc}
\usepackage{amsmath,amssymb,bbm,color}
\usepackage{graphicx}
\usepackage{epsfig}
\usepackage{euscript}
\usepackage{pslatex}
\begin{document}
\fontfamily{cmr}\selectfont

%
%
\title{\bf \fontfamily{cmr10}\selectfont  
Prediction for the Cosmological Constant and  Constraints on Susy GUTS in Resummed Quantum Gravity
}

\author{\fontfamily{cmr8}\selectfont  B.F.L. Ward\footnote{On Research Leave from Baylor University, Waco, TX, USA during 01/04/18  to 07/31/18  at Werner-Heisenberg-Institut, Max-Planck-Institut fuer Physik, Foehringer Ring 6, 80805 Muenchen, Germany}\\
Physics Department, Baylor University, One Bear Place \# 97316\\
Waco, Texas, 76798-7316, USA\\
bfl\_ward@baylor.edu\\
}

\maketitle

\vspace{-.25in}
\centerline{BU-HEPP-18-06}
\begin{abstract}
{\fontfamily{cmr8}\selectfont \baselineskip=10pt Working in the context of the Planck scale cosmology formulation of Bonanno and Reuter, we use our resummed quantum gravity approach to Einstein's general theory of relativity to estimate the value of the cosmological constant as $\rho_\Lambda =(0.0024 eV)^4$. We show that susy GUT models are constrained by the closeness of this estimate to experiment. We also address various consistency checks on the calculation. In particular, we use the Heisenberg uncertainty principle to remove a large part of the remaining uncertainty in our estimate of $\rho_\Lambda$.
}
\end{abstract}
\section{Introduction}
\label{intro}
The calculations in Refs.~\citen{reutera,laut,reuterb,reuter3,litim,perc} have given considerable support to Weinberg's suggestion~[\citen{wein1}] that the general theory of relativity may be asymptotically safe, with an S-matrix
that depends only on a finite number of observable parameters, due to
the presence of a non-trivial UV fixed point, with a finite dimensional critical surface
in the UV limit. The former authors,
using Wilsonian~[\citen{kgw}] field-space exact renormalization 
group methods, obtain results which support the existence of Weinberg's  UV fixed-point for the Einstein-Hilbert theory. 
Independently, in Refs.~\citen{bw1,bw2,bw2a,bw2i} we have shown that the extension of the amplitude-based, exact resummation theory of Refs.~\citen{yfs,jad-wrd} to the Einstein-Hilbert theory leads to UV-fixed-point behavior for the dimensionless
gravitational and cosmological constants. The attendant resummed theory, which we have called resummed quantum gravity, is actually UV finite.
In Refs.~\citen{ambj}, we note that causal dynamical triangulated lattice methods have been used to show more evidence for Weinberg's asymptotic safety behavior\footnote{At the expense of violating Lorentz invariance, the model in Ref.~\citen{horva} realizes many aspects
of the effective field theory implied by the anomalous dimension of 2 at the Weinberg
UV-fixed point.}.
\par
The results in Refs.~\citen{reutera,laut,reuterb,reuter3,litim,perc}, which are quite impressive, however, involve cut-offs and some dependence on gauge parameters
which remain in the results
to varying degrees even for products such as that for the UV limits of the 
dimensionless gravitational and cosmological constants.  Accordingly, we continue to
refer to the approach in 
Refs.~\citen{reutera,laut,reuterb,reuter3,litim,perc} as the 
'phenomenological' asymptotic safety approach.
Because the above noted 
dependencies are mild, the non-Gaussian UV 
fixed point found in these latter references is probably a physical result. 
But, until it is corroborated by a 
rigorously cut-off independent and gauge invariant calculation, the result cannot be considered final.
Such a calculation is possible to do in resummed quantum gravity. 
As the results from Refs.~\citen{ambj} involve 
lattice constant-type artifact 
issues, to be considered final, they too need to be corroborated by a rigorous calculation 
without such issues. Again, a possible answer is  resummed quantum gravity. Thus in what follows, we try to make contact with experiment on a stage that has been prepared for us.
\par
More specifically, the authors in  Refs.~\citen{reuter1,reuter2} have applied the attendant phenomenological
asymptotic safety approach in 
Refs.~\citen{reutera,laut,reuterb,reuter3,litim,perc} 
to quantum gravity 
to provide an inflatonless realization\footnote{The authors in Ref.~\citen{sola1} also proposed the attendant 
choice of the scale $k\sim 1/t$ used in Refs.~\citen{reuter1,reuter2}.} of the successful
inflationary model~[\citen{guth,linde}] of cosmology
: the standard Friedmann-Walker-Robertson classical descriptions 
are joined smoothly onto Planck scale cosmology developed from the attendant UV fixed point solution.
In this way, the horizon, flatness, entropy
and scale free spectrum problems are solved with quantum mechanical arguments.
\par 
The properties as used in Refs.~\citen{reuter1,reuter2} 
for the UV fixed point of quantum gravity are reproduced in Ref.~\citen{bw2i} using the new
resummed theory~[\citen{bw1,bw2,bw2a}] of quantum gravity with the bonus of 
``first principles''
predictions for the fixed point values of
the respective dimensionless gravitational and cosmological constants. 
In what follows, the analysis in Ref.~\citen{bw2i} is carried forward~[\citen{drkuniv}] to
an estimate
for the observed cosmological constant $\Lambda$ in the
context of the Planck scale cosmology of Refs.~\citen{reuter1,reuter2}.
We comment on the reliability of the result and present arguments~[\citen{eh-consist}] showing that the uncertainty of the estimate of  is at the level of
a factor of ${\cal O}(10)$, as the estimate will be seen
already to be relatively close to the observed value~[\citen{cosm1,pdg2008}]. The closeness to
the observed value of our estimate allows us to constrain susy GUT models
given that this closeness is now put on a more firm basis~[\citen{eh-consist}]. 
The closeness of our estimate to the experimental value again gives, at the least, some more credibility to the new resummed theory as well as to the methods in Refs.~\citen{reutera,laut,reuterb,reuter3,litim,perc,ambj}\footnote{We do want to continue to caution against overdoing this closeness to the experimental value.}.
\par
An important point of contact for our approach to quantum gravity is the pioneering result of Weinberg~[\citen{sw-sftgrav}] on summing soft gravitons. 
Weinberg showed that, in an on-shell $\alpha \rightarrow \beta$ process with transition rate $\Gamma^0_{\beta\alpha}$ without soft graviton effects, inclusion of the virtual soft
graviton effects results in the transition rate
\begin{equation}
\Gamma_{\beta\alpha} = \Gamma^0_{\beta\alpha} (\lambda/U)^B,
\end{equation}
where $\lambda$ is the infrared cutoff and $U$ is the Weinberg~[\citen{sw-sftgrav}] soft cutoff which is used define what is meant by infrared. Here, $B$ is given by
\begin{equation}
B=\frac{G_N}{2\pi}\sum_{n,m}\eta_n\eta_mm_nm_m\frac{1+\beta_{nm}^2}{\beta_{nm}(1-\beta_{nm}^2)^{1/2}}\ln\left(\frac{1+\beta_{nm}}{1-\beta_{nm}}\right)
\label{eq1I}
\end{equation}
where $G_N$ is Newton's constant, $\eta_n=+1 (-1)$ when particle $n$ is outgoing (incoming), respectively, and $\beta_{nm}$ is the relative velocity $$\beta_{nm}=\left[1-\frac{m_n^2m_m^2}{(p_np_m)^2}\right]^{1/2}$$ for particles $n$ and $m$ with masses $m_n$, $m_m$ and four momenta $p_n$, $p_m$, respectively. In the 2-to-2 case where 1 and 2 are incoming, 3 and 4 are outgoing, and all masses have the same value $m$, we see that (\ref{eq1I}) shows a growth of the damping represented by $B$
with large values of $U$ as the exponential of $-(4G_Ns/\pi)\ln2\ln(U/\lambda)$ for large values of the cms energy squared s for the wide-angle case with the scattering angle at $90^o$
in the center of momentum system. We will see in our discussion
below that we recover this same type of growth of the analog of B with large invariant squared masses in the context of resumming the large IR regime of quantum gravity.
\par  
We present the discussion as follows. 
In Section 2 we give a brief review of
the Planck scale cosmology presented phenomenologically
in Refs.~\citen{reuter1,reuter2}. 
In Section 3 we review our results in
Ref.~\citen{bw2i} for the dimensionless gravitational and cosmological constants
at the UV fixed point. In Section 4, using our results in Section 3 in the context of
the Planck scale cosmology 
scenario in Refs.~\citen{reuter1,reuter2}, we estimate 
the observed value of 
the cosmological constant $\Lambda$ and we use the attendant estimate to constrain
susy GUT's. We also address consistency checks on the analysis. Specifically, in Section 5, we use the consistency between the Heisenberg uncertainty principle and the
solutions of Einstein's equations to argue the that error on our estimate of $\Lambda$ is ${\cal O}(10)$. Section 6  contains our summary remarks.
\par
\section{\bf Review of Planck Scale Cosmology}
The Einstein-Hilbert 
theory with which we work is defined by the Lagrangian
\begin{equation}
{\cal L}(x) = \frac{1}{2\kappa^2}\sqrt{-g}\left( R -2\Lambda\right),
\label{lgwrld1a}
\end{equation} 
where $R$ is the curvature scalar, $g$ is the determinant of the metric
of space-time $g_{\mu\nu}$, $\Lambda$ is the cosmological
constant and $\kappa=\sqrt{8\pi G_N}$.  
The authors in Ref.~\citen{reuter1,reuter2}, using the phenomenological exact renormalization group
for the Wilsonian~[\citen{kgw}] coarse grained effective 
average action in field space,  
have argued that
the attendant running Newton constant $G_N(k)$ and running 
cosmological constant
$\Lambda(k)$ approach UV fixed points as $k$ goes to infinity
in the deep Euclidean regime. Accordingly, we have 
$k^2G_N(k)\rightarrow g_*,\; \Lambda(k)\rightarrow \lambda_*k^2$
for $k\rightarrow \infty$ in the Euclidean regime.\par
One may use a connection between 
the momentum scale $k$ characterizing the coarseness
of the Wilsonian graininess of the average effective action and the
cosmological time $t$ to make contact with cosmology. Using a phenomenological realization of this latter connection, the authors
in Refs.~\citen{reuter1,reuter2} 
show that the standard cosmological
equations admit of the following extension:
\begin{align}
(\frac{\dot{a}}{a})^2+\frac{K}{a^2}&=\frac{1}{3}\Lambda+\frac{8\pi}{3}G_N\rho,\cr
\dot{\rho}+3(1+\omega)\frac{\dot{a}}{a}\rho&=0,\;\cr
\dot{\Lambda}+8\pi\rho\dot{G_N}&=0,\;\cr
G_N(t)=G_N(k(t)),&\;
\Lambda(t)=\Lambda(k(t)).\cr
\label{coseqn1}
\end{align}
Here, $\rho$ is the density and $a(t)$ is the scale factor
with the Robertson-Walker metric representation given as
\begin{equation}
ds^2=dt^2-a(t)^2\left(\frac{dr^2}{1-Kr^2}+r^2(d\theta^2+\sin^2\theta d\phi^2)\right)
\label{metric1}
\end{equation}
where $K=0,1,-1$ correspond respectively to flat, spherical and
pseudo-spherical 3-spaces for constant time t.  
We take the equation of state   
$ 
p(t)=\omega \rho(t),
$
where $p$ is the pressure.
Proceeding phenomenologically, the attendant functional relationship between the respective
momentum scale $k$ and the cosmological time $t$ is determined
via
$
k(t)=\frac{\xi}{t}
$
for some positive constant $\xi$ determined
from constraints on
physically observable predictions.\par 
The authors in Refs.~\citen{reuter1,reuter2}, using the UV fixed points as discussed above for $k^2G_N(k)\equiv g_*$ and
$\Lambda(k)/k^2\equiv \lambda_*$ obtained from their phenomenological, exact renormalization
group (asymptotic safety) 
analysis, solve the cosmological system given above. For $K=0$, they find
a solution in the Planck regime where $0\le t\le t_{\text{class}}$, with
$t_{\text{class}}$ a ``few'' times the Planck time $t_{Pl}$, which joins
smoothly onto a solution in the classical regime, $t>t_{\text{class}}$,
which coincides with standard Friedmann-Robertson-Walker phenomenology
but with the horizon, flatness, scale free Harrison-Zeldovich spectrum,
and entropy problems all solved purely by Planck scale quantum physics.\par
The phenomenological nature of the analyses in Refs.~\citen{reuter1,reuter2} is made manifest
by  the dependencies of
the fixed-point results $g_*,\lambda_*$ on the cut-offs
used in the Wilsonian coarse-graining procedure, for example.
We point out that 
the key properties of $g_*,\; \lambda_*$ used for their analyses 
are that the two UV limits are both positive and that the product 
$g_*\lambda_*$ is only mildly cut-off/threshold function dependent.
In what follows, we review the predictions of resummed quantum gravity (RQG)~[\citen{bw1,bw2,bw2a}] in Refs.~\citen{bw2i} for these UV limits
and we show how to use the limits to predict~[\citen{drkuniv}] the current value of $\Lambda$.
For completeness, a brief review of the basic principles of RQG theory is given at the beginning of the next section. 
\par
\section{\bf Review of Resummed Quantum Gravity Prediction for $g_*$ and $\lambda_*$}
We start with the prediction for $g_*$ as it is presented in Refs.~\citen{drkuniv,bw2,bw2a,bw2i}. Here, we recapitulate
the main steps in the calculation.
\par
The graviton couples to an elementary particle 
in the infrared regime which we shall
resum independently of the particle's spin~[\citen{sw-sftgrav,wein-qft}]. This means that
we may develop the required calculational framework using a scalar field. We extend that framework
to spinning particles straightforwardly.
\par 
We start with the Lagrangian density for
the basic scalar-graviton system already given by Feynman in Refs.~\citen{rpf1,rpf2}:{\small
\begin{equation}
\begin{split}
{\cal L}(x) &= -\frac{1}{2\kappa^2} R \sqrt{-g}
            + \frac{1}{2}\left(g^{\mu\nu}\partial_\mu\varphi\partial_\nu\varphi - m_o^2\varphi^2\right)\sqrt{-g}\\
            &= \quad \frac{1}{2}\left\{ h^{\mu\nu,\lambda}\bar h_{\mu\nu,\lambda} - 2\eta^{\mu\mu'}\eta^{\lambda\lambda'}\bar{h}_{\mu_\lambda,\lambda'}\eta^{\sigma\sigma'}\bar{h}_{\mu'\sigma,\sigma'} \right\}\\
            & + \frac{1}{2}\left\{\varphi_{,\mu}\varphi^{,\mu}-m_o^2\varphi^2 \right\} -\kappa {h}^{\mu\nu}\left[\overline{\varphi_{,\mu}\varphi_{,\nu}}+\frac{1}{2}m_o^2\varphi^2\eta_{\mu\nu}\right]\\
            &  - \kappa^2 \left[ \frac{1}{2}h_{\lambda\rho}\bar{h}^{\rho\lambda}\left( \varphi_{,\mu}\varphi^{,\mu} - m_o^2\varphi^2 \right) - 2\eta_{\rho\rho'}h^{\mu\rho}\bar{h}^{\rho'\nu}\varphi_{,\mu}\varphi_{,\nu}\right] + \cdots\;.\\
\end{split}
\label{eq1-1}
\end{equation}}
Here,
$\varphi(x)$ can be identified as the physical Higgs field as
our representative scalar field for matter,
$\varphi(x)_{,\mu}\equiv \partial_\mu\varphi(x)$,
and $g_{\mu\nu}(x)=\eta_{\mu\nu}+2\kappa h_{\mu\nu}(x)$
where we follow Feynman and expand about Minkowski space
so that $\eta_{\mu\nu}=\text{diag}\{1,-1,-1,-1\}$.
We have introduced Feynman's notation
$$\bar y_{\mu\nu}\equiv \frac{1}{2}\left(y_{\mu\nu}+y_{\nu\mu}-\eta_{\mu\nu}{y_\rho}^\rho\right)$$ for any tensor $y_{\mu\nu}$\footnote{Our conventions for raising and lowering indices in the 
second line of (\ref{eq1-1}) are the same as those
in Ref.~\citen{rpf2}.}.
The bare(renormalized) scalar boson mass here is $m_o$($m$) 
and we set presently the small
observed~[\citen{cosm1,pdg2008}] value of the cosmological constant
to zero so that our quantum graviton, $h_{\mu\nu}$, has zero rest mass.
We return to the latter point, however, when we discuss phenomenology.
Feynman~[\citen{rpf1,rpf2}] has essentially worked out the Feynman rules for (\ref{eq1-1}), including the rule for the famous
Feynman-Faddeev-Popov~[\citen{rpf1,ffp1a,ffp1b}] ghost contribution required 
for unitarity with the fixing of the gauge
(we use the gauge of Feynman in Ref.~\citen{rpf1},
$\partial^\mu \bar h_{\nu\mu}=0$).
For more details of this material we refer to Refs.~\citen{rpf1,rpf2}. 
We turn now directly to the quantum loop corrections
in the theory in (\ref{eq1-1}).
\par
We have shown
\begin{figure}
\begin{center}
\includegraphics[width=80mm]{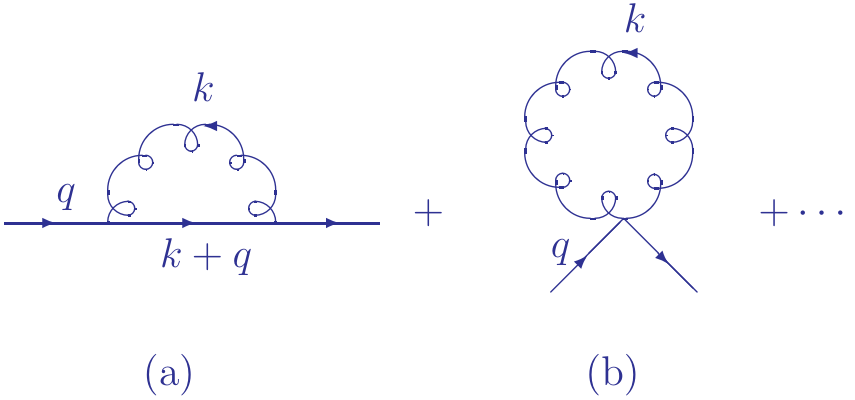}
\end{center}
\caption{\baselineskip=7mm     Graviton loop contributions to the
scalar propagator. $q$ is the 4-momentum of the scalar.}
\label{fig1}
\end{figure}
in Refs.~\citen{bw1,bw2,bw2a} that the large virtual IR effects
in the respective loop integrals for 
the scalar propagator in Fig.~\ref{fig1} in quantum general relativity  
can be resummed to the {\em exact} result
\begin{equation}
\begin{split} 
i\Delta'_F(k)&=\frac{i}{k^2-m^2-\Sigma_s(k)+i\epsilon}\\
&=  \frac{ie^{B''_g(k)}}{k^2-m^2-\Sigma'_s+i\epsilon}\\
&\equiv i\Delta'_F(k)|_{\text{resummed}}
\end{split}
\end{equation}
for{\small ~~~($\Delta =k^2 - m^2$)
\begin{equation}
\begin{split} 
B''_g(k)&= -2i\kappa^2k^4\frac{\int d^4\ell}{16\pi^4}\frac{1}{\ell^2-\lambda^2+i\epsilon}\\
&\qquad\frac{1}{(\ell^2+2\ell k+\Delta +i\epsilon)^2}\\
&=\frac{\kappa^2|k^2|}{8\pi^2}\ln\left(\frac{m^2}{m^2+|k^2|}\right).       
\end{split}
\label{yfs1} 
\end{equation}}
The latter form holds for the UV(deep Euclidean) regime, 
so that $\Delta'_F(k)|_{\text{resummed}}$ 
falls faster than any power of $|k^2|$ -- by Wick rotation, the identification
$-|k^2|\equiv k^2$ in the deep Euclidean regime gives 
immediate analytic continuation to the result in the last line of (\ref{yfs1})
when the usual $-i\epsilon,\; \epsilon\downarrow 0,$ is appended to $m^2$. See Ref.~\citen{bw1} for the analogous result
for $m=0$. Here, $-i\Sigma_s(k)$ is the 1PI scalar self-energy function
so that $i\Delta'_F(k)$ is the exact scalar propagator. As the residual $\Sigma'_s$ starts in ${\cal O}(\kappa^2)$,
we may drop it in calculating one-loop effects. 
When the respective analogs of $i\Delta'_F(k)|_{\text{resummed}}$\footnote{These follow from
the observation~[\citen{bw1,sw-sftgrav,wein-qft}] that the IR limit of the coupling of the graviton to a particle is independent of its spin.} are used for the
elementary particles, one-loop 
corrections are finite. In fact, the use of
our resummed propagators renders all quantum 
gravity loops UV finite~[\citen{bw1,bw2,bw2a}]. We have called the  attendant representation
of the quantum theory of general relativity the theory of
resummed quantum gravity (RQG).
\par
Specifically, we use our resummed propagator results, 
extended to all the particles
in the SM Lagrangian and to the graviton itself, working with the
complete theory
\begin{equation}
{\cal L}(x) = \frac{1}{2\kappa^2}\sqrt{-g} \left(R-2\Lambda\right)
            + \sqrt{-g} L^{\cal G}_{SM}(x)
\end{equation}
where $L^{\cal G}_{SM}(x)$ is SM Lagrangian written in diffeomorphism
invariant form as explained in Refs.~\citen{bw1,bw2a}, to show in the Refs.~\citen{bw1,bw2,bw2a} that the denominator for the propagation of transverse-traceless
modes of the graviton becomes ($M_{Pl}$ is the Planck mass)
\begin{equation}
q^2+\Sigma^T(q^2)+i\epsilon\cong q^2-q^4\frac{c_{2,eff}}{360\pi M_{Pl}^2},
\end{equation}
where we have defined
\begin{equation}
\begin{split}
c_{2,eff}&=\sum_{\text{SM particles j}}n_jI_2(\lambda_c(j))\\
         &\cong 2.56\times 10^4
\end{split}         
\end{equation}
with $I_2$ defined~[\citen{bw1,bw2,bw2a}]
by
\begin{equation}
I_2(\lambda_c) =\int^{\infty}_0dx x^3(1+x)^{-4-\lambda_c x}
\end{equation}
and with $\lambda_c(j)=\frac{2m_j^2}{\pi M_{Pl}^2}$ and
$n_j$ equal to the number of effective degrees of freedom~[\citen{bw1,bw2,bw2a}] of particle $j$. 
We refer the reader to Refs.~\citen{bw1} for the details of the derivation of the
numerical value of $c_{2,eff}$. In this way, we
identify (we use $G_N$ for $G_N(0)$) 
\begin{equation}
G_N(k)=G_N/(1+\frac{c_{2,eff}k^2}{360\pi M_{Pl}^2})
\end{equation}
and we compute the UV limit $g_*$ as
\begin{equation}
g_*=\lim_{k^2\rightarrow \infty}k^2G_N(k^2)=\frac{360\pi}{c_{2,eff}}\cong 0.0442.\end{equation}
\par
For the prediction for $\lambda_*$, we use the Euler-Lagrange
equations to get Einstein's equation as 
\begin{equation}
G_{\mu\nu}+\Lambda g_{\mu\nu}=-\kappa^2 T_{\mu\nu}
\label{eineq1}
\end{equation}
in a standard notation where $G_{\mu\nu}=R_{\mu\nu}-\frac{1}{2}Rg_{\mu\nu}$,
$R_{\mu\nu}$ is the contracted Riemann tensor, and
$T_{\mu\nu}$ is the energy-momentum tensor. Working then with
the representation $g_{\mu\nu}=\eta_{\mu\nu}+2\kappa h_{\mu\nu}$
for the flat Minkowski metric $\eta_{\mu\nu}=\text{diag}(1,-1,-1,-1)$
we see that to isolate $\Lambda$ in Einstein's 
equation (\ref{eineq1}) we may compute the
VEV (vacuum expectation value) of both sides of that equation. 
When we do this computation as described in Ref.~\citen{drkuniv}, we see that 
a scalar makes the contribution to $\Lambda$ given by\footnote{We note the
use here in the integrand of $2k_0^2$ rather than the $2(\vec{k}^2+m^2)$ in Ref.~\citen{bw2i}, to be
consistent with $\omega=-1$~[\citen{zeld}] for the vacuum stress-energy tensor.}
\begin{equation}
\begin{split}
\Lambda_s&=-8\pi G_N\frac{\int d^4k}{2(2\pi)^4}\frac{(2k_0^2)e^{-\lambda_c(k^2/(2m^2))\ln(k^2/m^2+1)}}{k^2+m^2}\cr
&\cong -8\pi G_N[\frac{1}{G_N^{2}64\rho^2}],
\end{split}
\label{lambscalar}
\end{equation} 
where $\rho=\ln\frac{2}{\lambda_c}$ and we have used the calculus
of Refs.~\citen{bw1,bw2,bw2a}.
The standard methods~[\citen{drkuniv}] 
then show that a Dirac fermion contributes $-4$ times $\Lambda_s$ to
$\Lambda$. The deep UV limit of $\Lambda$ then becomes, allowing $G_N(k)$
to run,
\begin{equation}
\begin{split}
\Lambda(k) &\operatornamewithlimits{\longrightarrow}_{k^2\rightarrow \infty} k^2\lambda_*,\;\\
\lambda_* &=-\frac{c_{2,eff}}{2880}\sum_{j}(-1)^{F_j}n_j/\rho_j^2\\
                 &\cong 0.0817
\end{split}
\end{equation} 
where $F_j$ is the fermion number of particle $j$, $n_j$ is the effective
number of degrees of freedom of $j$ and $\rho_j=\rho(\lambda_c(m_j))$.
In an exactly supersymmetric theory
$\lambda_*$ would vanish.\par
One may compare
the UV fixed-point calculated here, 
$$(g_*,\lambda_*)\cong (0.0442,0.0817),$$ with the estimate
$$(g_*,\lambda_*)\approx (0.27,0.36)$$
in Refs.~\citen{reuter1,reuter2}.
Here, one must keep in mind that 
the analyses in Refs.~\citen{reuter1,reuter2} did not include
the specific SM matter action and that there is definitely cut-off function
sensitivity to the results in the latter analyses. Qualitatively, the two sets of results are similar in that in both of them
$g_*$ and $\lambda_*$ are 
positive and are less than 1 in size.
Further discussion of the relationship between
our $\{g_*,\;\lambda_*\}$ predictions and those in Refs.~\citen{reuter1,reuter2} is given in Refs.~\citen{bw1}.
\par
\section{\bf Review of the RQG Estimate of $\Lambda$ and its Constraints on Susy GUTS}
When taken together with the results in Refs.~\citen{reuter1,reuter2}, the results given above allow us to estimate the value of $\Lambda$ today. We start from the normal-ordered form of Einstein's equation 
\begin{equation}
:G_{\mu\nu}:+\Lambda :g_{\mu\nu}:=-\kappa^2 :T_{\mu\nu}: .
\label{eineq2}
\end{equation}
If we use the coherent state representation of the thermal density matrix we can write
the Einstein equation in the form of thermally averaged quantities with
$\Lambda$ given by our result in (\ref{lambscalar}) summed over 
the degrees of freedom as specified above in lowest order. The Planck scale cosmology description of inflation in Ref.~\citen{reuter2}  gives the transition time between the Planck regime and the classical Friedmann-Robertson-Walker(FRW) regime as $t_{tr}\sim 25 t_{Pl}$. (We discuss in Ref.~\citen{drkuniv} the uncertainty of this choice of $t_{tr}$ and we present more on this uncertainty below.)
Starting with the quantity
\begin{equation}
\begin{split}
\rho_\Lambda(t_{tr}) \equiv\frac{\Lambda(t_{tr})}{8\pi G_N(t_{tr})}
         =\frac{-M_{Pl}^4(k_{tr})}{64}\sum_j\frac{(-1)^Fn_j}{\rho_j^2}
\end{split}
\label{eq-rho-lambda}
\end{equation}
and employing the arguments in Refs.~\citen{branch-zap} ($t_{eq}$ is the time of matter-radiation equality) we get the 
first principles field theoretic estimate
\begin{equation}
\begin{split}
\rho_\Lambda(t_0)&\cong \frac{-M_{Pl}^4(1+c_{2,eff}k_{tr}^2/(360\pi M_{Pl}^2))^2}{64}\sum_j\frac{(-1)^Fn_j}{\rho_j^2}\cr
          &\qquad\quad \times \frac{t_{tr}^2}{t_{eq}^2} \times (\frac{t_{eq}^{2/3}}{t_0^{2/3}})^3\cr
    &\cong \frac{-M_{Pl}^2(1.0362)^2(-9.194\times 10^{-3})}{64}\frac{(25)^2}{t_0^2}\cr
   &\cong (2.4\times 10^{-3}eV)^4.
\end{split}
\label{eq-rho-expt}
\end{equation}
Here, $t_0$ is the age of the universe and we take it to be $t_0\cong 13.7\times 10^9$ yrs. 
In the estimate in (\ref{eq-rho-expt}), the first factor in the second line comes from the radiation dominated period from
$t_{tr}$ to $t_{eq}$ and the second factor
comes from the matter dominated period from $t_{eq}$ to $t_0$ 
\footnote{The method of the operator field forces the vacuum energies to follow the same scaling as the non-vacuum excitations.}.
The estimate in (\ref{eq-rho-expt}) is close to the experimental result~[\citen{pdg2008}]\footnote{See also Ref.~\citen{sola2} for an analysis that suggests 
a value for $\rho_\Lambda(t_0)$ that is qualitatively similar to this experimental result.} 
$\rho_\Lambda(t_0)|_{\text{expt}}\cong ((2.37\pm 0.05)\times 10^{-3}eV)^4$. 
\par
We do believe our estimate 
of $\rho_\Lambda(t_0)$
represents some amount of progress in
the long effort to understand its observed value  
in relativistic quantum field theory. We do not consider the estimate to be a precision prediction,
as hitherto unseen degrees of freedom, such as a high scale GUT theory, 
may exist that have not been included in the calculation.\par
One may ask what would happen to our estimate if there were a GUT theory at high scales? The main
viable approaches in this regard involve susy GUT's.  For definiteness and purposes of illustration,
we will use the susy SO(10) GUT model in Ref.~\citen{ravi-1}
to illustrate how such a theory might affect our estimate of $\Lambda$.\par
In this model, the break-down of the GUT gauge symmetry to the 
low energy gauge symmetry occurs with an intermediate stage with gauge group
$SU_{2L}\times SU_{2R}\times U_1\times SU(3)^c$ where the final break-down to the Standard Model~[\citen{gsw,qcd}] gauge group, $SU_{2L}\times U_1\times SU(3)^c$, occurs at a scale $M_R\gtrsim 2TeV$ while the breakdown of global susy occurs at the (EW) scale $M_S$ which satisfies $M_R > M_S$. For our analysis
the key observation is that susy multiplets do not contribute to our formula
for $\rho_\Lambda(t_{tr})$ when susy is not broken -- there is exact cancellation between fermions and bosons in a given degenerate susy multiplet. Only the the broken susy multiplets can contribute. In the model at hand, these are just the multiplets associated with the known SM particles and the extra Higgs multiplet required by susy in the MSSM~[\citen{haber}].
In the light of recent LHC results~[\citen{lhc-susy}], we take for illustration the values $M_R\cong 4 M_S\sim 2.0{\text{TeV}}$ and set the following susy partner values:
\begin{equation}
\begin{split}
m_{\tilde{g}}&\cong 1.5(10){\text{TeV}},\;
m_{\tilde{G}}\cong 1.5{\text{TeV}},\;
m_{\tilde{q}}\cong 1.0{\text{TeV}},\;
m_{\tilde{\ell}}\cong 0.5{\text{TeV}},\;\\
m_{\tilde{\chi}^0_i}&\cong\begin{cases} &0.4{\text{TeV}},\;i=1\\
                                        & 0.5{\text{TeV}},\; i=2,3,4
                    \end{cases},\;\\
m_{\tilde{\chi}^{\pm}_i}&\cong  0.5{\text{TeV}},\; i=1,2,\;
m_{S} = .5{\text{TeV}},\; S=A^0,\; H^{\pm},\; H_2,
\end{split}
\end{equation}  
where we use a standard notation for the susy partners of the known quarks ($q\leftrightarrow \tilde{q}$), leptons ($\ell\leftrightarrow \tilde{\ell}$) and gluons ($G\leftrightarrow \tilde{G}$), and the EW gauge and Higgs bosons ($\gamma,\; Z^0,\; W^{\pm},\;H,$
$A^0,\;H^{\pm},\;H_2  \leftrightarrow \tilde{\chi}$)  with the extra Higgs particles denoted as usual~[\citen{haber}] by $A^0$(pseudo-scalar), $H^{\pm}$(charged) and $H_2$(heavy scalar). $\tilde{g}$ is the gravitino, for which we show two examples of its mass for illustration. 
From these particles we get the extra contribution 
\begin{equation}
\begin{split}
\Delta W_{\rho,\text{GUT}}=\sum_{j\in \{\text{MSSM low energy susy partners}\}}\frac{(-1)^Fn_j}{\rho_j^2}
          \cong 1.13(1.12)\times 10^{-2}
\end{split} 
\end{equation}
to the factor $W_\rho\equiv \sum_j\frac{(-1)^Fn_j}{\rho_j^2}$ on the RHS of 
our equation for $\rho_\Lambda(t_{tr})$ for the two respective values of $m_{\tilde g}$ indicated by the parentheses. The attendant values of $\rho_\Lambda$ are $-(1.67\times 10^{-3}\text{eV})^4(-(1.65\times 10^{-3}\text{eV})^4)$, respectively. Due to their signs, these results would appear to be in conflict with the positive observed value quoted above by many standard deviations. This last conclusion holds even when we allow for the considerable uncertainty in the various other factors, all positive,  multiplying $W_\rho$ in our formula for $\rho_\Lambda(t_{tr})$. To remedy this situation, we may either add new particles to the model, approach (A), or allow a near GUT scale soft susy breaking mass term for the gravitino, where the GUT scale
$M_{GUT}$ is $\sim 4\times 10^{16} GeV$ here~[\citen{ravi-1}], approach (B). Approach (A) doubles the number of quarks and leptons, but  inverts
the mass hierarchy between susy partners, so that the new squarks and sleptons are lighter than the new quarks and leptons. This also requires that
we increase $M_R,\; M_S$ so that we have the new quarks and leptons
at $M_{\text{High}}\sim 3.4(3.3)\times 10^3\text{TeV}$ while leaving their partners at $M_{\text{Low}}\sim .5{\text{TeV}}$. Approach (B) sets the mass of the gravitino soft breaking term to
$m_{\tilde{g}}\sim 2.3\times 10^{15}{\text{GeV}}$. What these results demonstrate is that our 
estimate in (\ref{eq-rho-expt}) can be used
as a constraint of general susy GUT models and we hope to explore such in more detail elsewhere. 
\par 
As we explain in Ref.~\citen{drkuniv},
we stress that we actually do not know the precise value of $t_{tr}$ at this point in the discussion to better than a couple of orders of magnitude. This translates to an uncertainty at the level of $10^4$ on
our estimate of $\rho_\Lambda$. We return to this issue in the next Section.
\par
We have not mentioned the effect of the various spontaneous symmetry vacuum energies on our 
$\rho_{\Lambda}$ estimate. From the standard methods we know for example that the energy of the broken vacuum for the EW case contributes an amount of order $M_W^4$ to $\rho_\Lambda$. If we consider the GUT 
symmetry breaking we expect an analogous contribution from spontaneous symmetry breaking of order $M_{GUT}^4$. The RHS of 
our equation for $\rho_\Lambda(t_{tr})$ is $$\sim (-(1.0362)^2W_\rho/64)M_{Pl}^4\simeq \frac{10^{-2}}{64}M_{Pl}^4.$$ It follows that including these broken symmetry vacuum energies would make relative changes in 
our results at the level of 
$$\frac{64}{10^{-2}}\frac{M_W^4}{M_{Pl}^4}\cong 1\times 10^{-65} $$ and $$\frac{64}{10^{-2}}\frac{M_{GUT}^4}{M_{Pl}^4}\cong 7\times 10^{-7},$$ respectively, where we use our value of $M_{GUT}$ given above in the latter 
evaluation for definiteness. We ignore
such small effects here.
\par
In discussing the impact of our approach to $\Lambda$
on the phenomenology of big bang nucleosynthesis(BBN)~[\citen{bbn}], we proceed as follows. We observe that the authors in Ref.~\citen{reuter2}
have noted that, when one passes from the Planck era to the FRW era,
a gauge transformation (from the attendant diffeomorphism invariance) is necessary to maintain consistency with
the solutions of the system (\ref{coseqn1})(or of its more general form as given below) at the transition time $t_{tr}$ at the boundary between the two regimes. From continuity of the Hubble parameter at $t_{tr}$ 
the authors in Ref.~\citen{reuter2} arrive at the gauge transformation on the
time for the FRW era relative to the Planck era $t\rightarrow t'=t-t_{as}$. It follows
that continuity of the Hubble parameter at the boundary gives
$$\frac{\alpha}{t_{tr}}=\frac{1}{2(t_{tr}-t_{as})}$$
when $a(t)\propto t^\alpha$ in the (sub-)Planck regime. This implies $$t_{as}=(1-\frac{1}{2\alpha})t_{tr}.$$ In our analysis, we have from Ref.~\citen{reuter2} the generic case $\alpha=25$, so that $t_{as}=0.98t_{tr}.$ Here, we use the diffeomorphism invariance of the theory to choose another coordinate transformation for the FRW era, namely, $$t\rightarrow t'=\gamma t$$ 
as a part of a dilatation
where $\gamma$ now satisfies the boundary condition required for continuity of the Hubble parameter at $t_{tr}$:
$$\frac{\alpha}{t_{tr}}=\frac{1}{2\gamma t_{tr}}$$  
so that $\gamma=\frac{1}{2\alpha}.$ According to the model in Ref.~\citen{reuter2}, for $t>t_{tr}$, one has the time $t'$ and an effective FRW cosmology with such a small value of $\Lambda$ that it may be treated as zero. Here, we extend this by retaining $\Lambda\ne 0$ so that we may estimate its value. With our 
diffeomorphism transformation between the (sub-)Planck regime and the FRW regime, we see that, at the time of BBN, the ratio of $\rho_\Lambda$ to
$\frac{3H^2}{8\pi G_N}$ is
\begin{equation} 
\begin{split} \Omega_\Lambda(t_{BBN}) &= \frac{M_{Pl}^2(1.0362)^29.194\times 10^{-3}(25)^2/(64 t_{BBN}^2)}{(3/(8\pi G_N))(1/(2\gamma t_{BBN})^2)}\cr &\cong \frac{\pi 10^{-2}}{24}\cr
&= 1.31\times 10^{-3}.\end{split}\label{bbneq1}\end{equation} At $t_{BBN}$ our $\rho_\Lambda$ has a negligible effect on the standard BBN phenomenology. 
Note that, in contrast to what happens in (\ref{eq-rho-expt}), the uncertainty in the value of $\alpha$ does not affect the estimate in (\ref{bbneq1}) because the factors of $\alpha^2=25^2$ cancel between the numerator and the denominator on the RHS in the first line of (\ref{bbneq1}).\par
Turning next to the issue of the covariance of the theory when $\Lambda$ and $G_N$ depend on time, Eqs.(\ref{coseqn1}) follow the corresponding realization of the improved Friedmann and Einstein equations as given in Eqs.(3.24) in Ref.~\citen{reuter1}.  
The more general
realization of (\ref{coseqn1}) is given in Eqs.(2.1) in Ref.~\citen{reuter2} and it is this latter realization which our discussions in this Section effectively followed. The two realizations differ in the solution of the Bianchi identity constraint:
$$D^\nu\left(\Lambda g_{\nu\mu}+8\pi G_N T_{\nu\mu}\right)=0;$$
for, this identity is solved in (\ref{coseqn1}) for a covariantly conserved
$T_{\mu\nu}$ whereas, in Eqs.(2.1) in Ref.~\citen{reuter2}, one has the modified conservation requirement
$$\dot{\rho}+3\frac{\dot{a}}{a}(1+\omega)\rho= -\frac{\dot{\Lambda}+8\pi\rho \dot{G}_N}{8\pi G_N}.$$
In (\ref{coseqn1}) the RHS of this latter equation is set to zero. The phenomenology from Ref.~\citen{reuter1} is qualitatively unchanged by the simplification in (\ref{coseqn1}) but the attendant details, such as the (sub-)Planck era exponent for the time dependence of $a$, etc., are affected, as is the relation between $\dot{\Lambda}$ and
$\dot{G}_N$ in (\ref{coseqn1}). We observe that (\ref{coseqn1}) contains a special case of the more general realization of the Bianchi identity requirement when both $\Lambda$ and $G_N$ depend on time and in this Section we use 
that more general realization. Note that covariant conservation of matter in the current universe is guaranteed only when $$\dot{\Lambda}+8\pi\rho \dot{G}_N=0$$ holds and that the case without such guaranteed conservation is possible provided the attendant deviation is small. See Refs.~\citen{bianref1,bianref2,bianref3}
for detailed studies
of such deviation, including its maximum possible size.\par
\section{Einstein-Heisenberg Consistency Condition and the Uncertainty of $t_{tr}$ }
Given the closeness of our estimate of $\rho_\Lambda$ to its observed value and the potential constraints on Beyond the Standard Model physics scenarios this closeness would obtain, it is appropriate to address the theoretical error of our estimate. This we do in this section using consistency arguments~[\citen{eh-consist}] based on the Heisenberg uncertainty principle and the solutions of Einstein's equation for the general theory of relativity.\par
Specifically, the basic physical idea which we wish to apply here is the known property of a de Sitter universe, which we describe here with the metric~[\citen{rn,ratra}] 
$$g_{\mu\nu}dx^{\mu}dx^{\nu}=dt^2-e^{2t/b}[dw^2+w^2(d\theta^2+\sin^2\theta d\phi^2)]$$ in an obvious notation, with $b=\sqrt{3/\Lambda}$
: if a light ray starts at the origin ($w=0$ here) and travels uninterruptedly, it never gets past the point $w=w_0\equiv b$ along its geodesic. Because we treat quantum mechanics as truly interwoven with the fabric of space-time, as it most certainly should be, according to Einstein's general theory of relativity, quantum mechanics must know about the latter limit for the quantum wave function of the photons in this light ray. According to the Heisenberg uncertainty principle, the uncertainty associated with the momentum conjugate variable to the coordinate distance $w$ is correspondingly bounded in the quantum theory of general relativity. To get a manifestation of the respective constraint, we use the results in Refs.~\citen{nactmn,duerr,cherni-tag,ratra} to 
check for the consistency of this bound with the effective scale $k$ associated to the running values of $G_N(k),\; \Lambda(k)$ as we discussed above.

More precisely, we start from the basic formulation of the Heisenberg uncertainty principle,
\begin{equation}
\Delta{p}\Delta{q}\ge \frac{1}{2},
\label{eheq1}
\end{equation}
where we define $\Delta{A}$ as the quantum mechanical uncertainty of the observable $A$ and $p$ is the momentum conjugate to the observable coordinate $q$. In our case, we have $q=w\cos\theta$ where $\theta$ is the polar angle when the direction of $\vec{k}$ is taken along the $\hat{z}$ direction and we may identify $\Delta{p}$ as our effective $k$, as $k$ represents the size of the mean squared momentum fluctuations in the universe that are effective for the running of the universe observables  $G_N(k),\; \Lambda(k)$. For the universe in the Planck regime, from the explicit solutions of the field equations in Refs.~\citen{nactmn,duerr,cherni-tag} we see that the solutions of the scalar field equations,\footnote{Spin continues to be an inessential complication here~[\citen{mlgbgr}].} in an appropriate set of coordinates, are spanned by plane waves in 3-space with Bessel/Hankel function-related dependence on time. We thus arrive at the estimate, at any given time, again using an obvious notation, 
\begin{equation}
(\Delta{q})^2\cong \frac{\int_{0}^{w_0}dw w^2 w^2<\cos^2\theta>}{\int_{0}^{w_0}dw w^2}=\frac{1}{5}w_0^2.
\label{eheq2}
\end{equation} 
From this estimate, we get the Einstein-Heisenberg consistency condition
\begin{equation}
k\ge \frac{\sqrt{5}}{2w_0}=\frac{\sqrt{5}}{2}\frac{1}{\sqrt{3/\Lambda(k)}}
\label{eheq3}
\end{equation}
where $\Lambda(k)$  follows from (\ref{eq-rho-expt}) (see  Eq.(52) in Ref.~\citen{drkuniv}):
\begin{equation}
\begin{split}
\Lambda(k)
         &=\frac{-\pi M_{Pl}^2(k)}{8}\sum_j\frac{(-1)^{F_j}n_j}{\rho_j^2}\cr
         &=\frac{-\pi M_{Pl}^2(1+c_{2,eff}k^2/(360\pi M_{Pl}^2))}{8}\sum_j\frac{(-1)^{F_j}n_j}{\rho_j^2}\cr
         &\cong \frac{\pi M_{Pl}^2(1+c_{2,eff}k^2/(360\pi M_{Pl}^2))\times 9.194\times 10^{-3}}{8}.
\end{split}
\label{eheq4}
\end{equation}
We argue that the Planck scale inflation must end when $k$ becomes too small to satisfy the condition (\ref{eheq3}). Accordingly, we estimate the transition time, $t_{\text tr}=\alpha/M_{Pl}=1/k_{\text{tr}}$, from the Planck scale inflationary regime to the Friedmann-Robertson-Walker regime via the value of $\alpha$ for which equality holds in (\ref{eheq3}). On our solving for $\alpha$ we get 
\begin{equation}
\alpha\cong 25.3,
\label{eheq5}
\end{equation}
which is in a agreement with the value $\alpha\cong 25$ implied by the numerical studies in Ref.~\citen{reuter1,reuter2}.\par
We conclude that the error on our estimate of $t_{\text tr}$ is at the level of a factor ${\cal O}(3)$ or less so that the uncertainty on our estimate of  $\Lambda$ is now reduced from a factor of 100~[\citen{drkuniv}] to a factor of ${\cal O}(10)$.\par
\section{Summary}
We have presented what amounts to a status report on the theory of resummed quantum gravity. We have reviewed the elements of the foundations of the theory, its predictions for the running of the Newton's constant and of the cosmological constant, and its prediction for the current value of the cosmological constant, where the latter prediction uses the Planck scale cosmology model of Bonanno and Reuter. Our prediction for the current value of the cosmological constant is close enough to experiment that it allows us to put constraints on possible susy GUT models, such as that presented in Ref.~\citen{ravi-1}. We have discussed as well the status of our RQG theory with respect to various other phenomenological and theoretical cosmological constraints. We have finally reviewed the use~[\citen{eh-consist}] of the Heisenberg uncertainty principle and the properties of the solutions of Einstein's equation for general relativity to constrain the main uncertainty in our estimate for the cosmological constant. In this way, we have reduced the latter uncertainty to a factor ${\cal O}(10)$. In principle, further improvement in this error should be possible.
We thank Profs. S. Bethke and W. Hollik for the support and kind
hospitality of the Werner-Heisenberg-Institut, MPI, Munich, Germany where a part of this work was done.\par
\par










\end{document}